\documentclass[twocolumm]{emulateapj}
\usepackage{xcolor}
\usepackage{natbib}
\usepackage{CJK}
\usepackage{hyperref}
\hypersetup{
colorlinks=true,
linkcolor=blue,
citecolor=blue}
\usepackage[all]{hypcap}

\newcommand{\ssst}{\scriptscriptstyle}
\newcommand{\E}[1]{\times 10^{#1}}
\newcommand{\etal}{et al.}
       
\newcommand{\RA}[3]{{#1}^{{\rm h}}{#2}^{{\rm m}}{#3}^{{\rm s}}}
\newcommand{\decl}[3]{{#1}^{\circ}{#2}'{#3}''}
\newcommand{\RAdot}[4]{{#1}^{{\rm h}}{#2}^{{\rm m}}{#3}\fs{#4}}
\newcommand{\decldot}[4]{{#1}^{\circ}{#2}'{#3}\farcs{#4}}

      \newcommand{\ps}{\,{\rm s}^{-1}}
\newcommand{\yr}{\,{\rm yr}}    \newcommand{\Msun}{M_{\odot}}
\newcommand{\cm}{\,{\rm cm}}    \newcommand{\km}{\,{\rm km}}
\newcommand{\kms}{$\km\ps$}
 \newcommand{\pc}{\,{\rm pc}}
        \newcommand{\K}{\,{\rm K}}

\newcommand{\Tmb}{T_{\rm mb}}

 \newcommand{\NHH}{N({\rm H}_{2})}
\newcommand{\VLSR}{V_{\ssst\rm LSR}}

\newcommand{\du}{d_{2.5}} 
    
\newcommand{\Rb}{R_{\rm b}} 
\newcommand{\Rm}{R_{\rm m}} 
\newcommand{\Vb}{V_{\rm b}}

\newcommand{\snr}{Tycho}

\newcommand{\twCO}{$^{12}$CO}   \newcommand{\thCO}{$^{13}$CO}

\newcommand{\Jotz}{$J$=1--0}    \newcommand{\Jtto}{$J$=2--1}
\newcommand{\Jttt}{$J$=3--2}
\newcommand{\vw}{v_{\rm w}}

\newcommand{\tw}{t_{\rm w}}

\newcommand{\caa}{Chinese Astron.\ Astrophys.\ }
\newcommand{\HI}{H{\footnotesize{I}}}

\slugcomment{Received 2016 March 21; 
accepted 2016 May 3; published 2016 July 18}

\begin{document}
\begin{CJK*}{UTF8}{bsmi}

\title{Expanding molecular bubble surrounding Tycho's supernova
remnant (SN~1572) observed with the IRAM 30~m telescope:
evidence for a single-degenerate progenitor}

\author{
Ping Zhou (周平) \altaffilmark{1,2},
Yang Chen (陳陽) \altaffilmark{1,2},
Zhi-Yu Zhang (張智昱)\altaffilmark{3,4},
Xiang-Dong Li (李向東)\altaffilmark{1,2},
Samar Safi-Harb\altaffilmark{5},
Xin Zhou (周鑫)\altaffilmark{6},
and Xiao Zhang (張瀟)\altaffilmark{1}
}

\altaffiltext{1}{Department of Astronomy, Nanjing University, 
Nanjing~210023, China; 
\href{mailto:pingzhou@nju.edu.cn}{pingzhou@nju.edu.cn}}
\altaffiltext{2}{Key Laboratory of Modern Astronomy and Astrophysics,
 Nanjing University, Ministry of Education, China}
\altaffiltext{3}{Institute for Astronomy, University of Edinburgh, 
Royal Observatory, Blackford Hill, Edinburgh EH9 3HJ, UK}
\altaffiltext{4}{ESO, Karl Schwarzschild Strasse 2, D-85748 Garching, 
Munich, Germany}
\altaffiltext{5}{Department of Physics and Astronomy, University of 
Manitoba, Winnipeg, MB R3T 2N2, Canada}
\altaffiltext{6}{Purple Mountain Observatory, CAS, 2 West Beijing Road, 
Nanjing 210008, China} 

\begin{abstract}

Whether the progenitors of SNe Ia are single-degenerate or 
double-degenerate white dwarf (WD) systems is a highly debated topic. 
To address the origin of the Type Ia Tycho's supernova remnant (SNR), SN 1572,
we have carried out a \twCO~\Jtto\ mapping and a 3-mm line survey
toward the remnant using the IRAM 30~m telescope.
We show that \snr\ is surrounded by a clumpy molecular bubble at a
local standard of rest velocity of $\sim 61~\km\ps$ which expands at 
a speed of $\sim 4.5$~\kms\ and has a mass of $\sim 220~\Msun$ (at the distance
of 2.5~kpc).
Enhanced \twCO~\Jtto\ line emission relative to \twCO~\Jotz\
emission and possible line broadenings (in velocity range $-64$ to $-60$~\kms) 
are found at the northeastern boundary of the SNR,
where the shell is deformed and decelerated.
These features, combined with the morphological correspondence between 
the expanding molecular bubble and \snr, suggest that the SNR is 
associated with the bubble at the velocity range $-66$ to $-57$~\kms.
The most plausible origin for the expanding bubble is the fast outflow 
(with velocity of hundreds \kms) driven from the vicinity of a 
WD as it accreted matter from a nondegenerate companion star.
The SNR has been expanding in the low-density wind-blown bubble, and the 
shock wave has just reached the molecular cavity wall.
This is the first unambiguous detection of an expanding bubble driven 
by the progenitor of a Type-Ia SNR,
which constitutes evidence for a single-degenerate progenitor for 
this SN Ia.

\end{abstract}

\section {Introduction}

Type Ia supernovae (SNe Ia) are catastrophic thermonuclear explosions
of massive carbon/oxygen white dwarfs (WDs) with mass exceeding a critical value.
Due to the low intrinsic scatter at the peak luminosity, SNe Ia
have been employed as ``standard candles'' for measuring distances
and determining cosmological parameters (Riess \etal\ 1998; Perlmutter 
\etal\ 1999).
The single-degenerate (SD) and double-degenerate (DD) models are 
the most popular scenarios for the SN Ia progenitors. 
The SD models assume a WD in a close binary system accreting material 
from a nondegenerate companion (Whelan \& Iben, 1973), while the DD 
models involve a merger of two WDs (Webbink, 1984).

The historical supernova remnant (SNR) \snr\ (or SN~1572, G120.1+1.4) 
is thought to originate from an SN Ia as evidenced by its 
light curve (Baade 1945; Ruiz-Lapuente 2004), X-ray spectroscopic
analysis (Badenes \etal\ 2006), and light echo spectrum (Krause 
\etal\ 2008).
However, there is no consensus on its progenitor system.
The SD scenario is supported by the discovery of a fast-moving, 
type G0--G2 subgiant (Tycho-G) in the SNR center, which was
suggested to be the surviving companion of the SN (Ruiz-Lapuente \etal\
2004).
Lu \etal\ (2011) found an X-ray arc inside \snr, 
which was probably produced by the interaction between the SN 
ejecta and the companion star's envelope lost due to the impact 
of the explosion.
The SD scenario has, however, been questioned (Badenes \etal\ 2007).
While there is a dispute over whether Tycho-G is the ex-companion
(see Xue \& Schaefer 2015 and references therein), other candidates 
such as Tycho-B (Kerzendorf \etal\ 2013)
and Tycho-E (Ihara \etal\ 2007) were also proposed.
Another independent method is needed to resolve the progenitor problem of \snr.

One key aspect of the SD and DD scenarios is the presence or absence,
respectively, of outflows from the binary system during the pre-SN 
evolution (Badenes \etal\ 2007).
According to the SD scenario, a strong wind is driven
to stabilize the mass transfer when the mass accretion rate exceeds 
a certain critical value (e.g., Hachisu \etal\ 1996; Li \& van den 
Heuvel 1997; Hachisu \etal\ 1999).
The strong wind could blow a low-density bubble before the SN explodes 
and the shock wave interacts with the dense material of the wind 
cavity (e.g, RCW~86; Vink \etal\ 1997; Broersen \etal\ 2014).
The size of the wind-blown bubble (WBB) depends not only on the wind
parameters, but also the on the density of the ambient medium
(e.g., Weaver \etal\ 1977; Koo \& Mckee 1992a, 1992b).

\capstartfalse
\begin{deluxetable*}{p{3.cm}ccccc}
\tabletypesize{\footnotesize}
\tablecaption{CO observations of \snr\ with the IRAM 30~m telescope}
\tablewidth{0pt}
\tablehead{
\colhead{Lines} & Position & Map Size & HPBW & $\NHH$ & Azimuth Angle$^a$\\
 & & & & $10^{20}~\cm^{-2}$}
\startdata
\twCO~\Jtto & ($\RAdot{00}{25}{21}{0}$, $\decldot{64}{08}{35}{0}$) 
& $12'\times 12'$ & $11''$ & - & -\\
\twCO~\thCO~\Jotz & P1($\RAdot{00}{25}{53}{1}$, $\decldot{64}{08}{24}{8}$) 
&  -  & $21''$ & 7.3 & $87\arcdeg$\\
\twCO~\thCO~\Jotz & P2($\RAdot{00}{25}{51}{7}$, $\decldot{64}{08}{55}{8}$) 
&  -  & $21''$ & 5.2 & $79\arcdeg$\\
\twCO~\thCO~\Jotz & P3($\RAdot{00}{25}{50}{5}$, $\decldot{64}{09}{23}{0}$) 
&  -  & $21''$ & 7.7 & $71\arcdeg$\\
\twCO~\thCO~\Jotz & P4($\RAdot{00}{25}{56}{0}$, $\decldot{64}{08}{46}{8}$) 
&  -  & $21''$ & 9.3 & $82\arcdeg$\\
\twCO~\thCO~\Jotz & P5($\RAdot{00}{25}{48}{8}$, $\decldot{64}{09}{51}{0}$) 
&  -  & $21''$ & 7.3 & $62\arcdeg$\\
\twCO~\thCO~\Jotz & P6($\RAdot{00}{25}{54}{9}$, $\decldot{64}{09}{17}{0}$) 
&  -  & $21''$ & 8.6 & $75\arcdeg$
\enddata
\label{T:obs}
  \tablenotetext{a}{\phantom{0}  The azimuth angle of P1-6 (counterclockwise
  from the north). The geometric center ($\RAdot{00}{25}{19}{4}$,
  $\decldot{64}{08}{13}{98}$, J2000) determined by Warren \etal\ (2005)
  is used here.
  }
\end{deluxetable*}
\capstarttrue

\snr\ is is proably the only known Type Ia SNR interacting with a 
molecular cloud (MC).
Radio observations spanning a 10 yr interval revealed slow expansion 
in the northeast, where the shell is encountering denser 
material (Reynoso \etal\ 1997, 1999).
Lee \etal\ (2004) found that the SNR is likely surrounded
by a molecular shell at $\VLSR \sim -67$ to $-60$~\kms\ with 
the Nobeyama \twCO~\Jotz\ observation (beam width of $16''$).
The MC in this velocity interval has also been 
studied in \twCO~\Jotz\ lines (beam width of $60''$; Cai \etal\ 2009), 
and in \twCO~\Jtto\ and \Jttt\ lines (beam width
of $130''$ and $80''$, respectively; Xu \etal\ 2011).
Shock interaction with the adjacent dense gas provides a 
reasonable mechanism for the hadronic $\gamma$-ray emission arising 
from \snr\ (Zhang \etal\ 2013).
The association of the MC with \snr\ has not been confirmed yet
and still requires kinematic evidence such as line broadening and
a large high-to-low excitation line ratio (see Jiang \etal\ 2010 
for observational evidence for judging the contact of SNRs with MCs).

To confirm or refute the SNR--MC association and to 
address the progenitor of \snr, we have performed high-resolution 
CO observations with the IRAM 30~m telescope.

\section{Observation}
We have performed observations of molecular lines toward \snr\ 
in 2013 August with the IRAM 30~m telescope.
The \twCO\ \Jtto\ (at 230.538~GHz) mapping was conducted 
with the on-the-fly position switching mode using the 9-pixel 
dual-polarization heterodyne receiver array (HERA) and the fast 
Fourier transform spectrometer (FTS).
The backend FTS provides a bandwidth of 0.5~GHz and a velocity resolution
of 0.063~\kms.
The \twCO~\Jtto\ observation covered a $12'\times 12'$ region centered
at ($\RA{00}{25}{21}$, $\decl{64}{08}{35}$, J2000), with an antenna half-power 
beam width (HPBW) of $10\farcs{7}$. 
The data were then converted to a cube with a spatial pixel size of $5\farcs3$.
After correcting the main-beam efficiency of 59\% at 230.5~GHz,
the mean rms noise level of the main-beam temperature ($\Tmb$) is 
0.32~K in each pixel with a velocity resolution of 0.5~\kms.

We carried out a line survey at the 3~mm band at six positions 
(labeled with plus signs in the bottom-right panel of 
\autoref{f:12co21_grid})
with the Eight Mixer Receiver (EMIR) of the IRAM 30~m telescope.
The backend FTS provided a 16-GHz frequency coverage in 92--100~GHz
and 108--115~GHz. 
The velocity resolution, angular resolution, and main beam efficiency
were about $0.5~\km\ps$, $21''$, and 78\%, respectively.
The mean rms noise levels of $\Tmb$ are 0.06--0.10~K at 115~GHz and
0.02--0.05~K at 110~GHz.
The information of the observed lines is summarized in \autoref{T:obs}.

In addition, we retrieved the public archival \thCO~\Jotz\ mapping data 
taken during 2004 May and June with the 13.7~m millimeter-wavelength 
telescope of the Purple Mountain Observatory (PMO) at Delingha in China.
The data with a velocity resolution of 0.11~\kms\ were convolved to an 
angular resolution of $2'$ (initial HPBW was $1'$) to match the 
grid spacing of $2'$.
We used the CLASS package of GILDAS to reduce the data and
the KARMA (Gooch 1996) and IDL software packages for 
the data analysis and visualization. 

\section{Results} \label{S:results}

\begin{figure*}[tbh!]
\centering
\includegraphics[angle=0, width=0.95\textwidth]{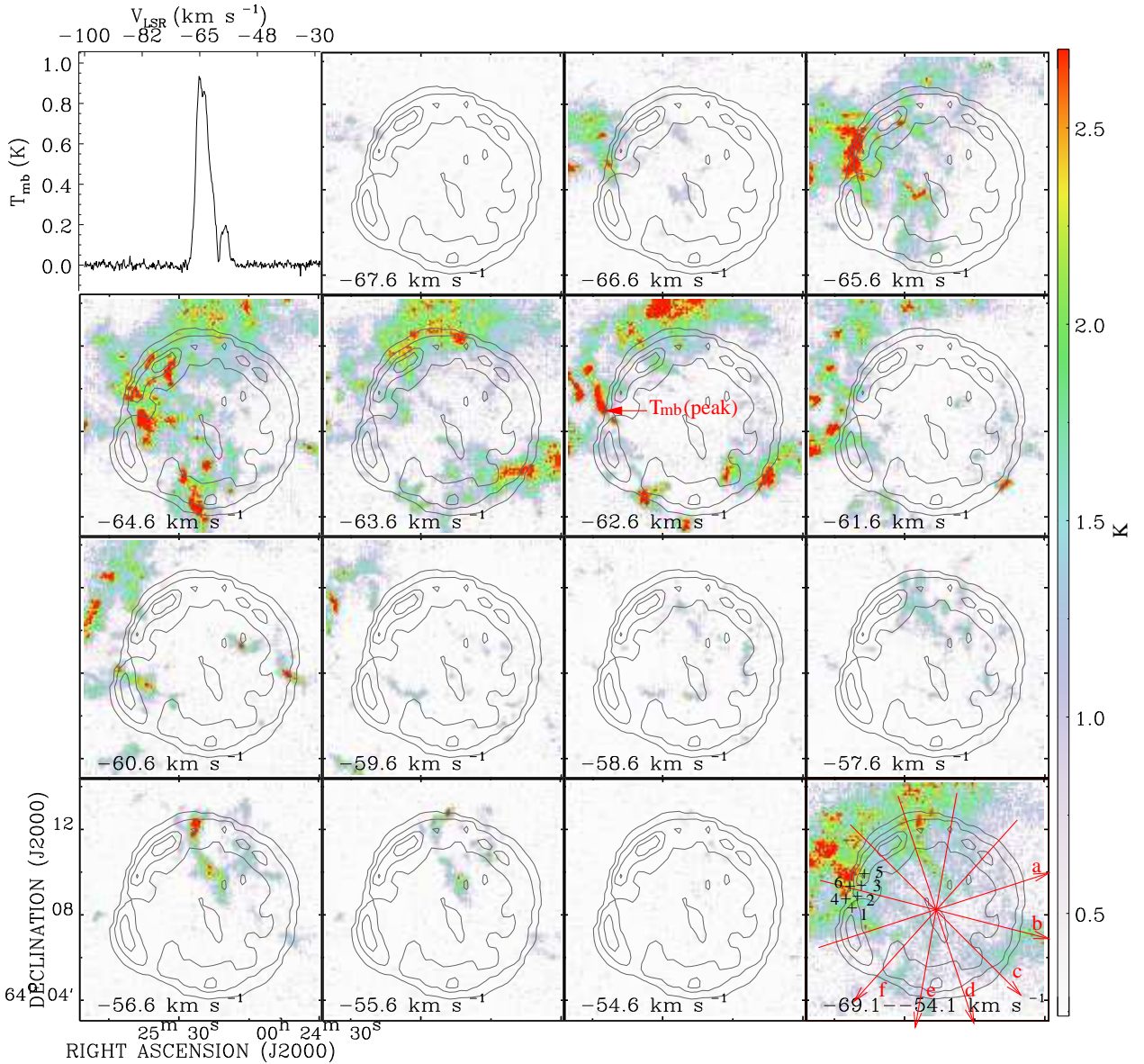}
\caption{
The upper-left panel shows the average spectrum of \twCO~\Jtto\
in the field of view.
The bottom-right panel shows the velocity-integrated intensity
in the velocity range $-68.1$--$-54.1$~\kms\ with the maximum 
intensity scale of 13.5~K.
The six plus signs label the positions where the 3-mm line survey
observations were made.
The data used in the position-velocity diagrams (in \autoref{f:pv}) 
are selected along the red lines.
Other panels reveal the \twCO\ \Jtto\ intensity channel maps 
with a step of 1~\kms, overlaid with contours 
of the 1.4~GHz continuum of the NRAO VLA Sky Survey.
The color bar shows the intensity scale of the channel maps.
}
\label{f:12co21_grid}
\end{figure*}

The field of \snr\ shows emission of \twCO~\Jtto\ in the LSR velocity range 
of $-67$ to $-56$~\kms\ (see \autoref{f:12co21_grid}).
The overall emission has two peaks around the $\VLSR=-65$~\kms\ 
and $-62$~\kms, with a few weak structures at the velocity 
range $-58$ to $-54~\km\ps$.
The molecular gas near \snr\ is highly clumpy, as revealed in the channel
maps.
The \twCO~\Jtto\ emission has a $\Tmb\lesssim 3~\K$ in most of 
the field, which is lower than the kinetic temperature of typical 
interstellar MC ($\sim 10$~K).
The low main-beam temperature further supports the MC being highly clumpy
(size $<0.14$~pc at an assumed distance of 2.5~kpc) and
not resolved in our observation.
The brightest \twCO~\Jtto\ emission ($\Tmb {\rm (peak)}=7.3~\K$) arises
from the southwestern end of a molecular strip 
($\RAdot{00}{25}{56}{96}$, $\decl{64}{08}{56}{77}$, J2000),
which in projection contacts the eastern concave of the radio shell 
(see \autoref{f:12co21_grid}).
At $\VLSR=-62.6\km\ps$, the relatively bright \twCO~\Jtto\ clumps
are generally distributed surrounding the shell of the SNR and display
similar morphology to that previously shown in the \twCO~\Jotz\ 
emission (Lee \etal\ 2004).
The clumps contacting the SNR in the east, southeast, and north are
consistent with the locations showing low expansion rate 
from the radio observations (Reynoso \etal\ 1997).

\begin{figure*}[tbh!]
\centering
\includegraphics[angle=0, width=0.95\textwidth]{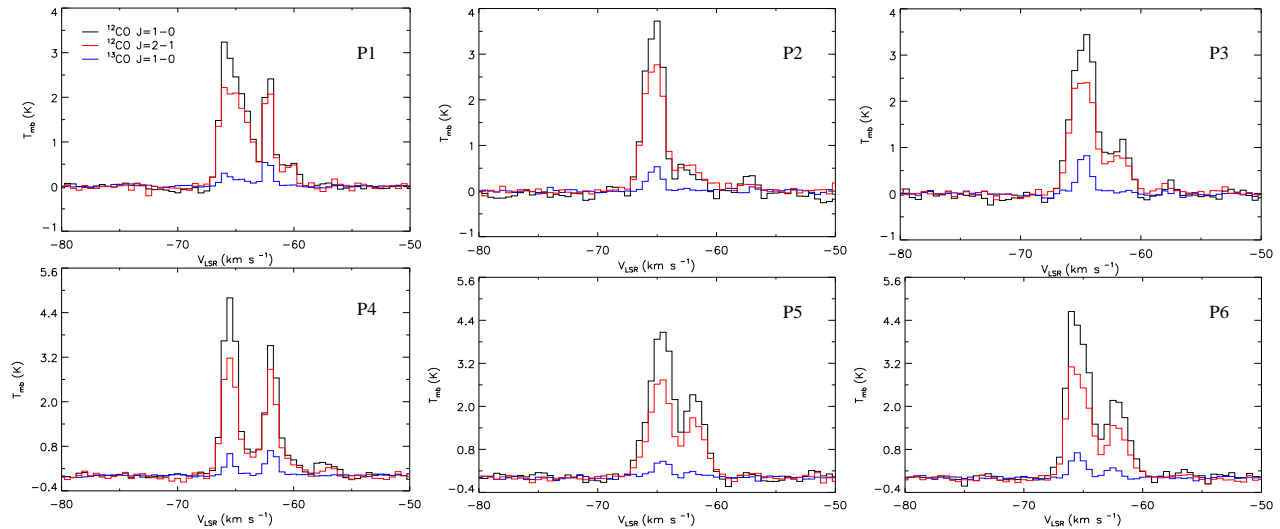}
\caption{
Spectra of CO and its isotope in six positions at the northeastern
boundary of \snr\ (see \autoref{f:12co21_grid} and \autoref{T:obs}).
}
\label{f:spec}
\end{figure*}

In order to search for kinematic evidence of the SNR--MC interaction, we
investigate the profiles of multiple molecular lines.
Only \twCO~\Jotz\ (115.271~GHz) and \thCO~\Jotz\ (110.201~GHz) lines 
were detected in the 3~mm band line survey toward the six positions at 
the northeastern boundary of \snr. 
\autoref{f:spec} shows the \twCO~\Jotz, \thCO~\Jotz, and \twCO~\Jtto\
lines at the six positions.
Here the \twCO~\Jtto\ data were convolved to $21\farcs{3}$ to match
the beam size of \twCO~\Jotz\ data.
We searched for broad \twCO\ line profiles in order to trace the kinematic 
structure of the shocked molecular gas, and high temperature of 
\twCO~\Jtto\ to trace heated gas.
At the positions ``P2'' and ``P3'', we find \twCO\ line wings at 
$\sim 63~\km\ps$ where \thCO\ emission was not detected. 
Such high \twCO/\thCO\ ratios suggest that the \twCO\ lines are 
optically thin due to low CO column density or/and large velocity 
gradient, as commonly seen in the line broadenings from the shocked clouds.
At position ``P2'', the \twCO~\Jtto\ emission in the velocity 
range $-64$ to $-60$~\kms\ is stronger than \twCO~\Jotz,
with a velocity-integrated intensity ratio of $\sim 1.6$.
Such a large high-to-low excitation line ratio needs high-excitation
conditions and
is generally regarded as evidence of shock heating and perturbation 
(Seta \etal\ 1998; Chen \etal\ 2014).
The azimuth angle of ``P2'' ($79 \arcdeg$; counterclockwise
from the north) is consistent with the angle where the forward shocks 
are significantly decelerated during 2003-2007 (see Fig.\ 5 in Katsuda 
\etal\ 2010).

\begin{figure*}[tbh!]
\centering
\includegraphics[angle=0, width=\textwidth]{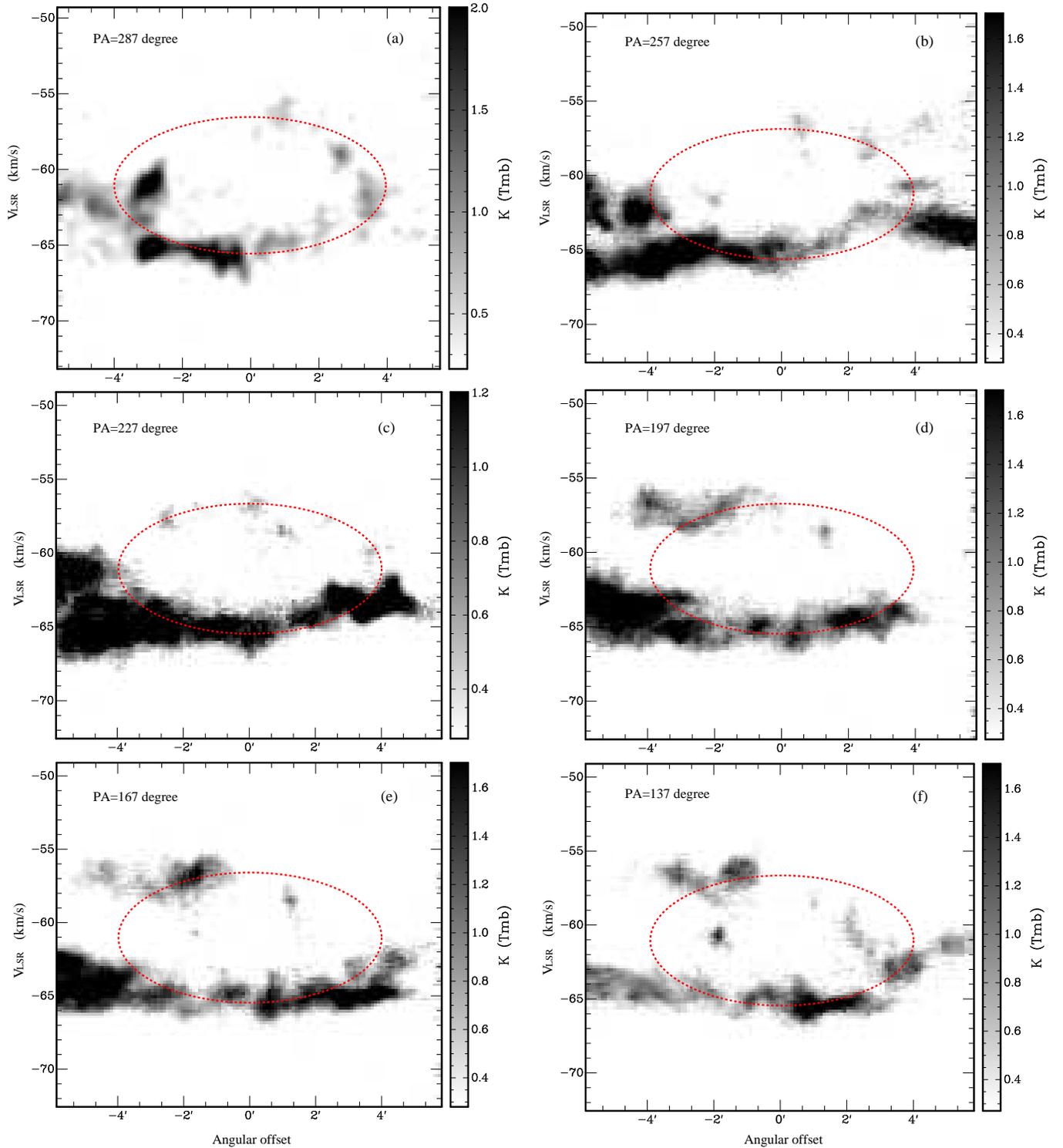}
\caption{
Position-velocity diagrams of \twCO~\Jtto\ emission across the SNRs
along six cuts (equivalent to optical long-slit spectra; 
the directions of the cuts with a $30\arcdeg$ decrement are 
displayed in the bottom-right panel of \autoref{f:12co21_grid}).
Panel a: the diagram is obtained along the $10\farcs{7}$-wide 
cut ``a'' with 
PA$=287\arcdeg$ and the data are smoothed with a Gaussian kernel of 
($16''$, $0.6~\km\ps$).
Panel b--f: the diagrams are for $2'$-wide cuts ``b''--``f''.
The data with signal-to-noise ratio $\ge 3$ are displayed.
The dashed ellipses indicate the sphere centered at ($\RA{00}{25}{20}$, 
$\decl{64}{08}{14}$, $61~\km\ps$), with a radius of $4'$, and expanding 
with a velocity of 4.5~\kms.
}
\label{f:pv}
\end{figure*}

An expanding bubble is revealed in the position-velocity diagrams of 
\twCO~\Jtto\ across the center of \snr\ (see \autoref{f:pv}).
The diagrams are obtained from six cuts starting from the 
position angle (PA) of $287\arcdeg$ and with a $30\arcdeg$ decrement 
(labeled with ``a''--``f'' in the bottom right panel of 
\autoref{f:12co21_grid}).
The morphology of the molecular gas revealed in the position-velocity 
dimensions is very similar to an expanding sphere centered at $\sim 61~\km\ps$
with an angular size of $\sim 8'$ and a velocity of $\sim 4.5~\km\ps$.
Most of the gas is at the blue-shifted side ($\VLSR=-66$ to $-60$~\kms).
The consistency of the size and position of the expanding circular gas 
with those of \snr, together with the cavity morphology 
(see \autoref{f:12co21_grid}) and large \twCO~\Jtto-to-\Jotz\ 
ratio at $\VLSR\sim -64$ to $-60~\km\ps$ (see
\autoref{f:spec}), implies that \snr\ is associated
with the expanding molecular bubble at $\VLSR=-66$ to $-57$~\kms.

\section{Discussion}

\subsection{Properties of the MC surrounding \snr}

Using the IRAM 30~m observations, we show that \snr\ is 
evolving in an expanding molecular bubble at $\VLSR\sim 61~\km\ps$.
The shock of the SNR might not have extensively impacted the 
molecular gas, but only contacted partially, e.g., in the northeast where the 
rim is deformed and decelerated as revealed in radio (Reynoso \etal\ 1997) 
and X-ray observations (Katsuda \etal\ 2010).
Although the data do not show very broad CO lines even with the best angular 
resolution in CO observations obtained to date ($10\farcs7$),
the large \twCO~\Jtto-to-\Jotz\ ratio and \twCO-to-\thCO\ ratios
are present in the northeast of the remnant.
This can be interpreted as the shocked clumps remaining unresolved
with current observations (low brightness temperature; see
Section~\ref{S:results}), and the shock front having just reached the 
molecular medium in the recent past (within a few tens of years; 
Reynoso \etal\ 1999) and swept up a small amount of gas.

\begin{figure}[tbh!]
\centering
\includegraphics[angle=0, scale=2, width=0.48\textwidth]{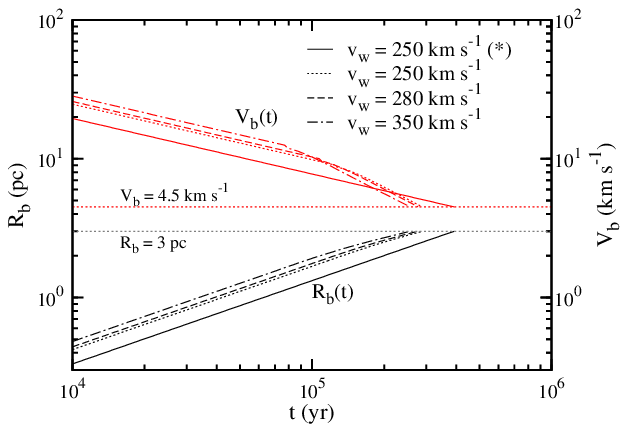}
\caption{
Exemplified possible evolutionary paths of the wind bubble with 
various sets of parameters. Paths A--D (with parameters tabulated in 
\autoref{tab:para}) are plotted with solid, dotted, dashed, and 
dot-dashed lines, respectively. 
The star sign indicates the evolution path that the wind lasts until 
the SN explosion (path A).
}
\label{f:evol}
\end{figure}

This molecular gas can be the dense target of collision by the 
cosmic-ray protons accelerated by Tycho's blast shock, which gives 
rise to $\gamma$-ray emission due to hadronic interaction.
A high ambient gas density provides a possibility of the relatively 
low fraction of energy deposited to the shock-accelerated protons 
in the young Tycho SNR (Zhang \etal\ 2013).

The suggested distance of \snr\ varies between 1.7 and 5~kpc according to 
different methods 
(Albinson \etal\ 1986; Schwarz \etal\ 1995; Krause \etal\ 2008; Hayato \etal\ 2010).
As suggested in Zhang \etal\ (2013; see also Lee \etal\ 2004), the system 
velocity of $-62~\km\ps$ of the MC allows the alternative distances of 4.0 
and 2.5~kpc, and the 
absence of an \HI\ absorption feature at $-46$ to $-41$~\kms\ 
(Tian \& Leahy 2011) is suggestive of the nearer distance.
Hence, we adopt a distance of $\sim 2.5$~kpc for \snr.

The column density of the molecular gas at the six positions at the 
northeastern boundary of the SNR is estimated with the \thCO\ line
detected at $\VLSR=-67$ to $-60$~\kms.
Assuming that the rotational levels of the \thCO\ molecules are in local 
thermodynamic equilibrium and the \thCO\ lines are optically thin,
we obtain a column density of H$_2$ 
$\NHH=1.21\E{20}\int T_{\rm mb,13CO}(v)dv [1-\exp(-5.29/T_{\rm ex})]^{-1}
\cm^{-2}$, where the relative abundance ratio of H$_2$ to \thCO\ is taken
as $5\E{5}$ (Dickman 1978).
A typical temperature of 10 K in MCs is adopted here for the
excitation temperature $T_{\rm ex}$.
The $\NHH$ values at the six positions at the northeastern boundary 
of the \snr\ are derived to be in the range (5.2--$9.3)\E{20}~\cm^{-2}$ 
as tabulated in \autoref{T:obs}.

We use the PMO \thCO~\Jotz\ line mapping to estimate the mass of
the MC surrounding the SNR. 
The \thCO\ emission is selected in a circular region with a radius 
of $6'$ and centered at point
($\RAdot{00}{25}{19}{4}$, $\decldot{64}{08}{13}{98}$, J2000).
The region encloses bright \twCO~\Jtto\ emission present in
\autoref{f:12co21_grid}.
The mean $\NHH$ in the region is $\sim 1.6\E{20}~\cm^{-2}$ and
the mass of the MC is $\sim 220~\du^2~\Msun$, where
$\du$ is the distance in units of 2.5~kpc.
This mass is one order of magnitude lower than that
derived with \twCO\ lines and the CO-to-H$_2$ conversion factor
(Cai \etal\ 2009). 
The standard conversion factor is a rough estimate of the 
global mass of a giant MC with a mass of $10^4$--$10^6~\Msun$ 
(Dame \etal\ 2001), and has a large variability on small 
spatial scales (Bolatto et al. 2013).

\subsection{Expanding molecular bubble as an implication for 
a SD progenitor}

The bubble surrounding \snr\ has an inner radius of about $4'$ 
(corresponding to $3~\du$~pc) and a mass of $\sim 220~\du^2~\Msun$.
Such an expanding bubble could be a WBB created with
a central star/system, while the shock wave of the SNR is not a plausible
source since the shock wave has just reached the boundary
of the molecular cavity and the velocity width (FWHM) of the
CO lines in the field is smaller than the expanding velocity 
of $5~\km\ps$.

First, we can exclude a massive star as the central source.
If the molecular bubble here was created by a massive star,
using the linear relationship between the maximum 
WBB sizes of main-sequence OB stars in the molecular environment 
$R_{\rm m}$ and the initial masses of the stars $M$ (Chen \etal\ 2013):
$p_5^{1/3}R_{\rm m} (\pc)\approx 1.22M/\Msun - 9.16$,
where $p_5$ is the interclump pressure $p/k$ scaled to $10^5~\cm^{-3} \K$
($p_5\sim 1$ in the interclump medium of MCs; Chevalier 1999; Blitz 1993; 
Krumholz \etal\ 2009),
the mass of the star would be $\gtrsim 10~\Msun$ (B2 type or earlier).
Such massive early-type stars have not been detected inside the bubble.

The alternative explanation is related to the progenitor system of \snr.
Substantial outflow with a mass-loss rate $\dot{M}\sim 10^{-6}~\Msun\yr^{-1}$ 
has been suggested to occur when a carbon/oxygen WD accretes
matter from a main-sequence or (slightly) evolved companion star. 
The wind may have a velocity $\vw$ larger than a few hundred \kms 
(Kato \& Hachisu 1999).

The fast wind from the accretion disk can evacuate a low-density cavity 
in which radiative losses are not important (Badenes \etal\ 2007; 
Koo \& Mckee 1992a,b).
The structure of the interaction of a strong wind with the interstellar medium
is described by the four zones:
(1) freely expanding wind; (2) shocked stellar wind; (3) a shell
of shocked interstellar gas; and (4) ambient gas (Weaver \etal\ 1977).
The expanding molecular bubble surrounding \snr\ has an inner radius 
$\Rb\sim 3~\du$~pc and an expansion velocity $\Vb\sim 4.5~\km\ps$.
In the adiabatic condition, if the wind lasts until the SN explosion
(e.g, see Fig.\ 1e in Han \& Podsiadlowski 2004),
the radius of the bubble (corresponding to the outer boundary of the shocked wind)
is $\Rb=0.66(\dot{M} \vw^2/\rho_0)^{1/5} \tw^{3/5}$,
where $\rho_0=1.4 n_0 m_{\rm H}$ is the mass density of the ambient gas,
with $n_0$ the number density of H atoms (Weaver \etal\ 1977).
The duration of the wind is $\tw=(3/5) \Rb/\Vb \sim 3.9\E{5}~\du\yr$.
The density $n_0\sim18~\du^{-1}~\cm^{-3}$ is estimated from the mass, 
$220~\du^2~\Msun$, of the swept-up molecular gas surrounding the
progenitor of \snr.
We estimate the wind velocity as $\vw\sim 250~ (\dot{M}/3\E{-7} \Msun 
\yr^{-1})^{-1/2}~\du^{1/2}\km\ps$
(see evolutionary path A in \autoref{tab:para} for the case with 
$\dot{M}=3\E{-7} \Msun \yr^{-1}$).
Here we constrain the wind velocity to be $\ga250\km\ps$ so that the
modeled bubble is in the energy conservation stage.
In this stage, the density of adiabatic gas in the shocked wind
has a flat density profile (see Eq.\ 16 in Weaver \etal\ 1977), which 
could be consistent with the suggestion that the SNR is evolving in 
a uniform medium (Badenes \etal\ 2007; Yamaguchi \etal\ 2014).
Since the mass-loss rates and durations of the wind vary in 
different outflow models (Han \& Podsiadlowski 2004), if the wind stops 
long before the SN explosion, the observed $\Rb$ and $\Vb$ values can give
multiple solutions of wind parameters
\footnote{If the wind terminates at $t_t$ (when the bubble size was $R_t$ and 
velocity was $V_t$) prior to the SN explosion, we assume the bubble will 
continue to expand driven by the thermal pressure ($p$) of the shocked wind region.
From the mass equation $dM=4 \pi R_b^2 \rho_0 dR_b$
and the momentum equation $d(M\Vb)= 4\pi \Rb^2 p$, the bubble radius is govern by
$d\Rb/dt=\sqrt{V_t^2R_t^6+2a_0(\Rb-R_t)}/\Rb^3$, where $a_0=3E_0 R_t^2/(2\pi\rho_0)$
with $E_0=(5/11)\dot{M}\vw^2 t_t/2$ (Weaver et al.\ 1977).
}.
Some exemplified evolutionary paths of the bubble are shown in \autoref{f:evol}
with the parameters listed in \autoref{tab:para}, 
where a typical $\dot{M}$ value of $10^{-6}\, M_{\odot}\yr^{-1}$ is adopted.
The gas density at the center of the shocked wind region is given by 
Castor et al. (1975)
$n_b~=~0.01~( n_0/1\ {\rm cm^{-3}} )^{19/35}
 ( \dot{M}/10^{-6}\,M_{\odot}\,{\rm yr}^{-1} )^{6/35}$
 $( \vw/2000\, {\rm km\, s^{-1}} )^{12/35}
 ( \tw/10^6\ {\rm yr} )^{-22/35} 
 \ \cm^{-3}$,
which is further scaled as $n_b(t_b)\sim n_b(t_t) (R_t/\Rb)^3$
if the elapse time $t_{elapse}=t_b-t_t$ between the wind stop and SN explosion cannot be ignored.
The density estimates, $n_b(t_b)\sim0.02$--$0.04\cm^{-3}$, for the interior of the bubble
are consistent with the previous constraints for the SNR preshock gas
obtained from X-ray ($\lesssim 0.2~\cm^{-3}$; Katsuda \etal\ 2010)
and infrared measurements ($\sim0.03$--$0.5~\cm^{-3}$, Williams 
\etal\ 2013; the observed elevated density in the northeast can be 
related to the denser gas near the cavity wall).
All of the wind parameters $\vw$, $\tw(t_t)$ and $\dot{M}$ match 
those expected in the SD scenario.
There is also some observational evidence to support the existence of
recurrent novae in outbursts before the SN explosion 
(Hachisu \& Kato 2001), 
which may be an alternative outflow source to affect the circumstellar 
medium of the SNR.

\capstartfalse
\begin{center}
\setlength{\tabcolsep}{1.5pt} 
\begin{deluxetable}{lccccccc}
\tablecaption{Parameters for the exemplified evolutionary paths of the bubble.}
\tabletypesize{\footnotesize}
\tablehead{
\colhead{Path} & $v_{\rm w}$  & $\dot{M}$ & $t_t$ & $t_{elapse}$  & $n_{\rm b}$($t_t$)  & $n_{\rm b}$($t_{\rm b}$)  &\\
      & (km s$^{-1}$) & ($10^{-6} M_{\odot} \yr^{-1}$)& ($10^{5}$ yr) & ($10^{5}$ yr) & (cm$^{-3}$)& (cm$^{-3}$)&}
\startdata
  A & 250 & 0.3 & 3.9 & 0   & 0.04 & 0.04 & \\
  B & 250 & 1.0 & 1.3 & 1.6 & 0.09 & 0.02 & \\
  C & 280 & 1.0 & 1.1 & 1.7 & 0.10 & 0.03 & \\
  D & 350 & 1.0 & 0.8 & 1.8 & 0.13 & 0.02 &
\enddata
\label{tab:para}
\end{deluxetable}
\end{center}
\capstarttrue
\vspace{0.1in}

\vspace{-0.5in}
A WBB will reach a maximum size $\Rm$ when it becomes
in pressure equilibrium with the ambient medium:
$\Rm=[(1/2)M_{\rm_w}\vw^2/(2\pi p)]^{1/3}$, where
$M_{\rm w}$ is the total mass-loss of the wind 
(Chevalier 1999).
For this expanding bubble, we have $\Rm>\Rb=3\du~\pc$, and hence
we set a constraint on the mean wind velocity:
$\vw> 120 (M_{\rm w}/0.5~\Msun)^{-1/2} p_5^{1/2}~\du^{3/2}~\km\ps$.
Our results support the theoretical predictions of a strong wind
(over $100~\km\ps$) driven by the SD progenitor system.

An earlier wind from the post-AGB star in the planetary nebula
phase is unlikely to explain the current molecular bubble.
Assuming that the bubble is created by an ejection of mass ($\sim 1~\Msun$
at a velocity of $\sim 10^3~\km\ps$ for planetary nebulae) before the
formation of a WD, the slowly expanding molecular bubble should have 
been in the radiative evolution phase with an age (Chevalier 1974): 
$0.3 \Rb/\Vb\sim 2\E{5}~\yr$, which is much smaller than the duration 
of the later evolution of either SD
($\sim 10^6~\yr$; Hachisu \etal\ 1996) or DD progenitor systems
($>10^6~\yr$; Kashi \& Soker 2011).

The disk wind and the later mass transfer should not last longer than the 
lifetime of the MCs (of the order of 10~Myr; 
see Mac Low \& Klessen 2004 and references therein), or the molecular
bubble would have vanished.
Actually, the time between the onset of the wind and the SN explosion 
is less than a few Myr according to various SD models 
(see Badenes \etal\ 2007 and references therein).
Compared with the core-collapse SNe, SNe Ia have much longer delay time 
(elapsed time between the formation of the star and the explosion SN; 
$>40$~Myr; Maoz \etal\ 2011).
The MC surrounding \snr\ was probably newly formed in the vicinity of 
the progenitor binary, rather than the natal cloud.

\section{Summary}

We have performed a \twCO~\Jtto\ mapping and a 3-mm line survey 
toward the Type Ia SNR \snr\ using the IRAM 30~m telescope.
The main results are summarized as follows:

\begin{enumerate}
\item

SNR \snr\ appears to be confined in the clumpy ($<11''$) molecular gas at 
$\VLSR\sim -64$ to $-60$~\kms.
In this velocity range, we have found a large \twCO~\Jtto-to-\Jotz\ ratio 
($\sim 1.6$) and large \twCO\ to\thCO\ ratios at the northeast of the 
remnant, where the shell is deformed and decelerated.
These features provide kinematic evidence for the interaction between Tycho 
and the molecular gas.

\item We discover the expanding motion ($\sim 4.5~\km\ps$) of the 
molecular bubble at $\VLSR= -61~\km\ps$.
The mass of the molecular bubble is $\sim 220~\du^2\Msun$ and the 
mean column density is $\sim 1.6\E{20}~\cm^{-2}$.
The SNR has been expanding in the low-density WBB
and the shock wave has just reached the molecular cavity wall. 

\item
The most plausible origin for the expanding bubble is the fast outflow (with 
velocity hundreds of \kms) driven from the vicinity of the 
progenitor WD as it accreted matter from a nondegenerate companion star. 
The shocked wind confined in the molecular bubble has a
uniform density ($\sim 0.02$-$0.04\cm^{-3}$).

\item
The MC surrounding \snr\ was probably newly formed in the vicinity of 
the progenitor binary, rather than the natal cloud.

\item 
This is the first unambiguous detection of the expanding bubble driven by 
the progenitor of the Type Ia SNR, which provides important and 
independent evidence to support that the progenitor of \snr\ was an SD system.

\end{enumerate}

\begin{acknowledgements}
P.Z.\ is thankful to Dr.\ Fang-Jun Lu, Dr.\ Zhi-Yuan Li, and Prof.\ Noam 
Soker for helpful discussions and comments.
P.Z.\ and Y.C.\ thank the support of NSFC grants 11503008, 11233001
and 11590781, the 973 Program grant 2015CB857100, 
and grant 20120091110048 by the Educational Ministry of China.
Z.Z.\ acknowledges support from the European Research Council (ERC)
in the form of an Advanced Grant, COSMICISM.
X.L.\ was supported by NSFC under grant numbers 11133001 and 11333004.
S.S.H.\ acknowledges support from the Canada Research Chairs program,
the NSERC Discovery Grants program, and the Canadian Space Agency.
\end{acknowledgements}

\end{CJK*}

\end{document}